# Image Deraining and Denoising Convolutional Neural Network For Autonomous Driving

Kaige Wang*
New Technology Development Department
Institute of Microelectronics of the Chinese Academy of Sciences; UCAS
Beijing, China
wangkaige@ime.ac.cn

Long Chen
Intelligent Manufacturing Electronics Research Center
Institute of Microelectronics of the Chinese Academy of Sciences
Beijing, China
chenlong@ime.ac.cn

Tianming Wang
Intelligent Manufacturing Electronics Research Center
Institute of Microelectronics of the Chinese Academy of Sciences
Beijing, China
wangtianming@ime.ac.cn

Qixiang Meng
New Technology Development Department
Institute of Microelectronics of the Chinese Academy of Sciences
Beijing, China
mengqixiang@ime.ac.cn

Huatao Jiang
New Technology Development Department
Institute of Microelectronics of the Chinese Academy of Sciences
Beijing, China
jianghuatao@ime.ac.cn

Lin Chang*
New Technology Development Department
Institute of Microelectronics of the Chinese Academy of Sciences
Beijing, China
changlin@ime.ac.cn

*Abstract*—Perception plays an important role in reliable decision-making for autonomous vehicles. Over the last ten years, huge advances have been made in the field of perception. However, perception in extreme weather conditions is still a difficult problem, especially in rainy weather conditions. In order to improve the detection effect of road objects in rainy environments, we analyze the physical characteristics of the rain layer and propose a deraining and denoising convolutional neural network structure. Based on this network structure, we design ablation experiments and experiment results show that our method can effectively improve the accuracy of object detection in rainy conditions.

## I. INTRODUCTION

Perception is critical for autonomous vehicles. It needs accurate, real-time and efficient environmental awareness in all-weather driving conditions. However, the current research is mainly focused on object detection under ideal weather conditions. Perception in severe weather (e.g., rain, snow, fog, etc.) is still need to be developed, while rainy days are common in daily life. The generation of raindrops and rain fog as a complex atmospheric process will cause different types of visibility degradation [1]. Generally, raindrops or rain bands in the vicinity tend to obstruct or distort the content of the background scene, while rain bands in the distance tend to produce atmospheric occlusion effects such as fog and clouds, which blur the image content [2][3][4], thereby adversely affecting the robustness of environment perception tasks, such as image segmentation [5], object detection [6], pattern recognition [7], etc.

The traditional method of computer vision to remove rain from images is the model-driven method, which use a priori knowledge of the physical properties of rainwater and background scenes to perform deraining optimization [8]. Typical techniques include morphological component analysis [9], non-local mean filtering [10], sparse cod deraining [11] and Gaussian mixture model (GMM) [12]. These methods have high computational complexity, long processing time and insignificant rain removal effect. In recent years, the emergence and development of convolutional neural network methods have provided a data-driven deraining method which is different from model-driven rain removal. The core idea of the data-driven deraining method is to set a specific network structure and loss function based on a large number of training image pairs in an end-to-end manner, and adjust the network structure parameters through backpropagation algorithm [13], so as to achieve the deraining function using the trained neural network.

Fu et al. [14][15] proposed the earliest deep learning-based method, which uses the principle of CNN to remove raindrops from the high-frequency part of the rainy image. Garg and Nayar et al. [16] made early attempts to remove rainwater in video, and proposed that by directly increasing the exposure time or reducing the depth of field of the camera, the impact of rain can be reduced or even eliminated without changing the appearance of the scene in the video [17]. Different from the above method, Zhang et al. [18] introduced an adversarial learning method to enhance the authenticity of the rain-removing pictures. Wei et al. [19] proposed a semi-supervised training model, which can be better generalized to practical tasks. Zong et al. [20] proposed a semi-supervised video rain removal network, which uses a dynamic generator to fit the rain layer in order to characterize the deep characteristics of the rain layer better. Although the above-mentioned deep learning-based deraining methods have achieved some results, due to the lack of quantity and quality of the dataset and the lack of in-depth analysis of the rainfall model, the existing methods can be improved in practical applications.

In this paper, we propose a deraining and denoising convolutional neural network, which draws on the residual networks [21] and the semi-supervised video deraining networks. Besides, the network introduces noise analysis to achieve good results on the synthetic rain dataset. We also verify that the deraining and denoising network plays a positive role in improving the performance of object detection in rainy environment.

The research contents and innovations of this paper are specified as follows.

Firstly, an end-to-end image deraining and denoising convolutional neural network is proposed, including a raindrop feature extraction layer and a random noise extraction layer, where the raindrop feature extraction layer uses residual modules and convolutional modules, etc., and the random noise layer uses cross-layer connections, convolutional modules and residual modules, etc. The training set of the network contains different synthetic rain datasets and some real rain datasets, and data enhancement is performed to make the network generalize better.

Secondly, the deraining and denoising images generated by the deraining and denoising network are input to the object detection network YOLO v3 to test its effect on the object detection performance, and to analyze the reasons for the differences in object detection effect.

The remainder of the paper is organized as follows. In Section II, we describe the composition of the dataset and the structure of the neural network. We present and analyze the results of the deraining and denoising experiments and the object detection experiment in Section III. Finally, Section VI concludes the research work of this paper.

## II. Deraining Network

### A. Dataset

The training dataset mainly includes the synthetic rainfall dataset, as well as a small part of the real world dataset.

The synthetic dataset includes rain and no rain images automatically generated by the software, including Rain100L, Rain100H and other commonly used baseline datasets in the field of rain removal. The synthetic dataset can provide a large number of comparison images such as Fig. 1, which will help the network to fully learn the rain removal methods. The real dataset includes rain and no rain images at the same place at the same time, including manual shooting and fixed-point camera images such as Fig. 2. The real dataset makes the convolutional neural network learn real raindrop information and improve its generalization performance.

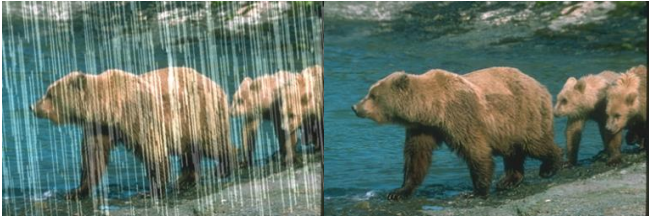

Fig. 1. Synthetic photos.

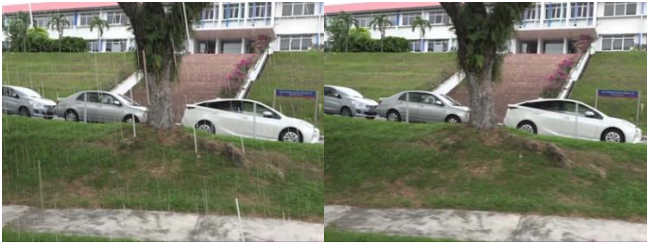

Fig. 2. Real world photos.

### B. The architecture of the network

During rainfall, the size of raindrops is generally about 0.1-3.5 mm, and during the falling process, their shape will change dramatically by gravity, air resistance, and wind, which will produce random occlusion of the target in the image, while Grag [21] et al. found that spherical or ellipsoidal raindrops are like lenses that refract or reflect light, and the pixel values near the raindrops are usually brighter than background pixels. Besides, in the actual imaging process, random noise is generated on the final acquired image due to the influence of the shooting environment light, temperature and exposure time of the shooting device.

In this paper, based on the analysis of the rainfall image process above, it is concluded that the rainfall images consisting of the following three parts, as shown in Eq. 1.

$$Y = X + R + Noise \qquad (1)$$

$Y$ represents the observed picture with rain, $X$ represents the picture without rain, $R$ represents the rainfall layer, and $Noise$ represents the random noise caused by rainfall. From the above Eq. 1, if we want to get the no rain picture $X$ according to the observed rain picture $Y$, we should calculate or approximate the rainfall layer $R$ and the random noise $Noise$, which is calculated by Eq. 2.

$$X = Y - R - Noise \qquad (2)$$

In the actual calculation process, it is difficult to get the mathematical expressions of rainfall layer and random noise directly, and the shape characteristics of the two layers in the image are obviously different, so the two layers are approximated as two different high-frequency components. So both of the model parameters can be approximated by training the convolution and residual networks, and then the clear image can be recovered from the rainy image.

The network structure is shown in Fig. 3 below.

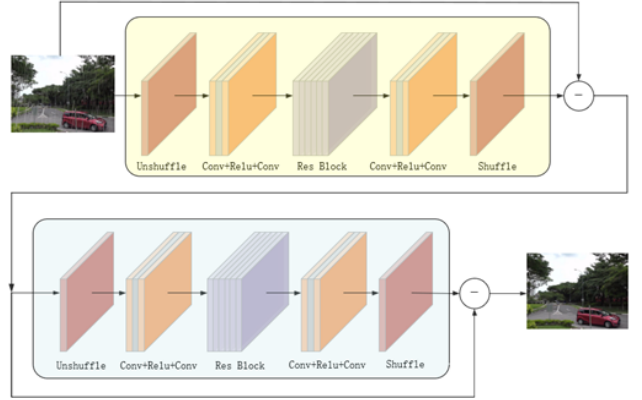

Fig. 3. Our Deraining and Denoising network structure.

We proposed a hierarchical cascaded neural network structure. The upper network structure is based on residual blocks, Pixel-Unshuffle and Pixel-Shuffle layers are introduced to accelerate the computation, which are mainly used to extract the rainfall layer $R$. The rainfall image is subtracted from the rainfall layer to obtain the rain-free image $Y - R$ with the random noise layer superimposed, and then the lower network structure is used to extract the random noise layer $Noise$. Although the structure of the upper and lower layers is similar, the internal parameters of the convolution layer, shuffle layer and residual module are different, so they can play the role of extracting the rainfall and random noise layers respectively.

The structure of the residual block in the network structure model diagram is shown in Fig. 4. $x$ denotes the input and $F(x)$ denotes the output of the residual block before the activation function in the second layer. σ denotes the activation function Relu, and if there is no jump connection, the residual block is a normal two-layer network. The network in the residual block uses a convolutional layer. The final output of this residual block $H(x)$ is shown in Eq. 3.

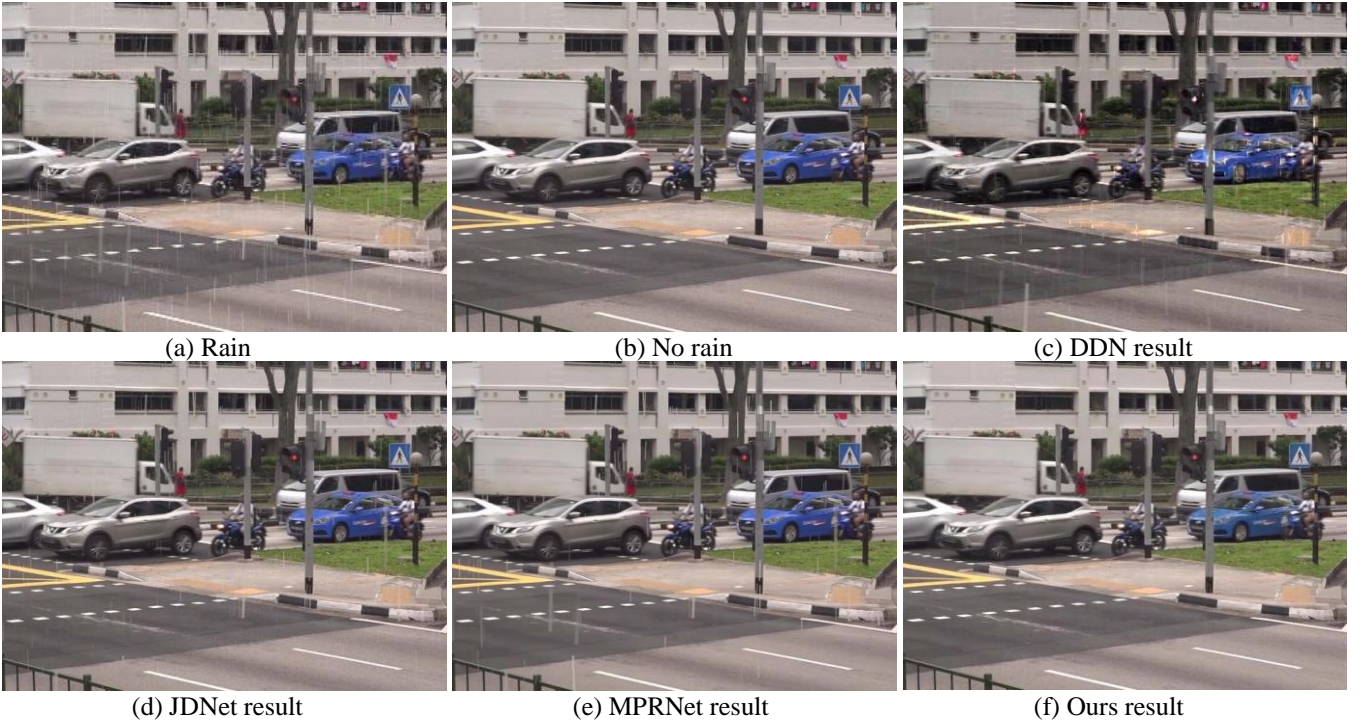

(a) Rain      (b) No rain      (c) DDN result

(d) JDNet result      (e) MPRNet result      (f) Ours result

Fig. 5 Rain, No rain, DDN result, JDNet result, MPRNet result, Ours result, separately.

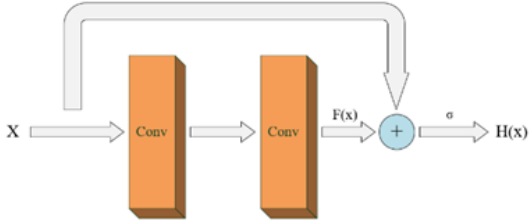

Fig. 4. Residual block structure.

$$H(x) = \sigma(F(x) + x) \quad (3)$$

### C. Design of loss function

The Mean Square Error (MSE) function is selected as the loss function to update the neural network parameters in this paper, and its calculation formula is shown in Eq. 4.

$$Loss = \frac{1}{n}\sum_{i=1}^{m}|Y^i - X^i|^2 \quad (4)$$

In Eq. 4, $Loss$ denotes the network training error, and $n$ denotes the number of images per batch, $m$ denotes the number of image pixel points, $X^i$ represents the value of each pixel point of the rain-free image, and $Y^i$ represents the value of each pixel point of the deraining and denoising image after the neural network processing.

## III. EXPERIMENTAL RESULTS

In order to prove the effectiveness of the rainfall model proposed in this paper, we selected some synthetic rain pictures on urban roads, and tested the rain removal effect through visual sense and rain removal metrics. Subsequently, in order to test whether the image deraining and denoising has an effect on object detection, we designed a set of ablation experiments to further explore the role of the deraining and denoising convolutional neural network. The results of the above experiments are as follows.

### A. Deraining experimental results

In terms of network rain and noise removal effect as shown in Fig. 5, our network has a certain removal effect for raindrops, especially the smaller raindrops, and a certain weakening effect for larger raindrops. Our proposed network outperforms all other networks in terms of visual perception.

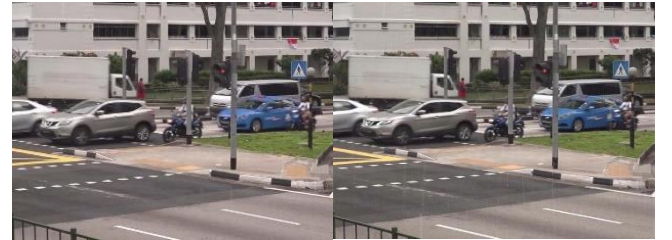

(a) Only Deraining      (b) Only Denoising

Fig. 6 Only Deraining and Only Denoising result.

Fig. 6 shows the result of image with only deraining and only denoising. It can be seen that the result of only deraining operation still contains visible raindrops, and the result of only denoising operation has more raindrops. The result of deraining and denoising Fig. 5(f) is better than Fig. 6(a) and Fig. 6(b) in terms of visual perception.

We also compared the rain removal metrics PSNR and SSIM with other classic networks in terms of evaluation indicators. PSNR is usually used to evaluate the quality of the processed image compared with the reference image. Its algorithm principle is based on the loss between the corresponding pixel points. The calculation method is shown in Eq. 5 and Eq. 6.

$$MSE = \frac{1}{col*row}\sum_{i=1}^{col}\sum_{j=1}^{row}\bigl(X(i,j) - Y(i,j)\bigr)^2 \quad (5)$$

$$PSNR = 10 * \lg\left(\frac{(2^n-1)}{MSE}\right) \quad (6)$$

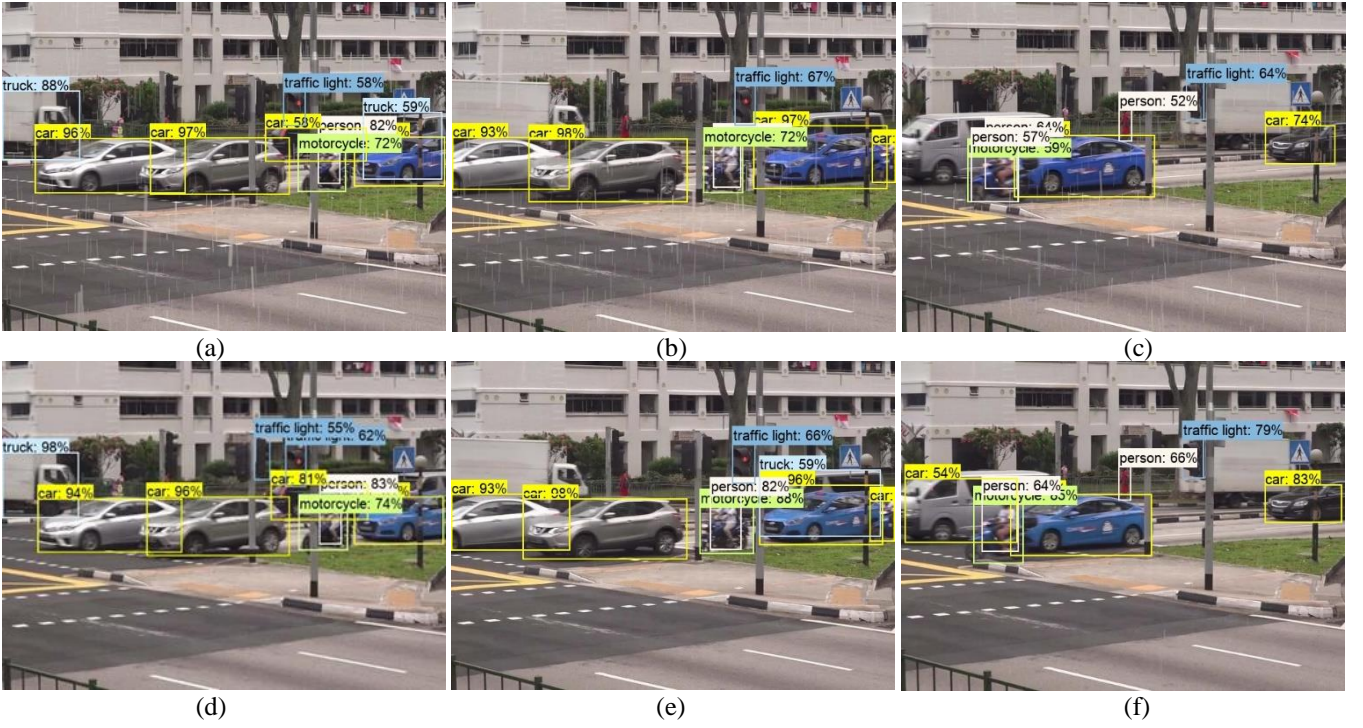

Fig. 7 Object detection results for rainy images and our deraining and denoising images. Figure (a), Figure(b) and Figure (c) are rainy images, Figure (d), Figure(e) and Figure (f) are our deraining and denoising images.

SSIM stands for structural similarity, which allows a more direct comparison between the structure of the deraining image and the reference image. The calculation of SSIM is mainly compared from three aspects: brightness, contrast comparison and structure. The SSIM similarity calculation between deraining image and no rain image is shown in Eq. 7.

$$SSIM(x, y) = [l(x, y)]^\alpha [c(x, y)]^\beta [s(x, y)]^\gamma \quad (7)$$

The PSNR and SSIM metrics of each network calculated by the above equations are shown in Table I and Table II.

TABLE I. PSNR METRICS COMPARISON

| PSNR Metrics | Rain PSNR | DDN PSNR | JDNet PSNR | MPRNet PSNR | Ours PSNR |
|---|---|---|---|---|---|
| image1 | 30.50 | 22.25 | 32.07 | 31.71 | 36.57 |
| image2 | 30.04 | 22.15 | 34.32 | 33.55 | 37.73 |
| image3 | 30.94 | 22.25 | 32.42 | 32.03 | 37.69 |
| image4 | 31.04 | 22.20 | 34.30 | 33.95 | 37.62 |
| image5 | 30.79 | 22.21 | 31.43 | 30.97 | 37.14 |
| Avg100. | 30.80 | 22.22 | 33.04 | 32.65 | 37.52 |

TABLE II. SSIM METRICS COMPARISON

| SSIM Metrics | Rain SSIM | DDN SSIM | JDNet SSIM | MPRNet SSIM | Ours SSIM |
|---|---|---|---|---|---|
| image1 | 0.9284 | 0.8343 | 0.9545 | 0.9519 | 0.9705 |
| image2 | 0.9283 | 0.8417 | 0.9635 | 0.9588 | 0.9755 |
| image3 | 0.9294 | 0.8381 | 0.9536 | 0.9484 | 0.9747 |
| image4 | 0.9285 | 0.8377 | 0.9637 | 0.9612 | 0.9754 |
| image5 | 0.9317 | 0.8389 | 0.9568 | 0.9499 | 0.9746 |
| Avg100. | 0.9299 | 0.8376 | 0.9590 | 0.9555 | 0.9745 |

TABLE III. PSNR METRICS COMPARISON

| PSNR Metrics | Only Deraining | Only Denoising | Both |
|---|---|---|---|
| Avg100. | 34.37 | 33.87 | 37.52 |

TABLE IV. SSIM METRICS COMPARISON

| SSIM Metrics | Only Deraining | Only Denoising | Both |
|---|---|---|---|
| Avg100. | 0.9688 | 0.9558 | 0.9745 |

It can be seen from Table I and Table II that compared with the original rain image and the excellent network results in recent years, our network has achieved better results in key metrics. PSNR is improved by 14.36% compared with MPRNet and 12.89% compared with JDNet network. SSIM is closer to the ideal value 1 than other networks.

It can be seen from Table III and Table IV that the metrics of deraining and denoising images are significantly better than deraining or denoising only.

B. *Object detection results*

We selected a synthetic road video under rainy weather from the test set, then split this rainy video into image frames and sent it to the YOLO v3 object detection network. At the same time, we split this rainy video into image frames and input them into the deraining and denoising neural network described in this paper to remove the rain and noise. After deraining and denoising is completed, they are sent to the same YOLO v3 object detection network for object detection.

As shown in the above Fig. 7, our deraining and denoising network is effective for the object detection task in rainy conditions. The details are as follows.

*1) Image deraining and denoising is helpful in increasing the object detection rate in the road:* It can be seen from figure 7(a) and figure 7(d) that the number of identified traffic lights has increased in comparison with the rain image after object detection. This phenomenon was prevalent in the tests on several hundred images.

*2) Image deraining and denoising can improve the confidence level of the identified road objects:* From figure 7(b) and figure 7(e), it can be found that the confidence level of road objects in the deraining and denoising images is generally improved compared to the images with rain, and this phenomenon was prevalent in the tests on several hundred images.

*3) Image deraining and denoising can improve the accuracy of target categories:* From figure 7(c) and figure 7(f), it can be found that the object detection after deraining and denoising images compared to the images with rain can correct the error of the object detection network identifying the motorcycle as human. In the test with several hundred images, this happened rarely as the object detection network is heavily pre-trained.

The comparison above demonstrates that the robustness of the deraining and denoising images is stronger compared to the images with rain, which helps to enhance the robustness of object detection in autonomous driving.

## IV. CONCLUSIONS

In order to improve the image quality in rainy conditions, and improve the robustness of the object detection task, this paper proposes a deraining and denoising convolutional neural network for perception of automatic driving. The network draws on the residual modules and the semi-supervised video deraining network, separates the rain layer and noise layer from rainy image through the convolutional layer, residual layer and shuffle layer, etc. By training on RTX 3090 after 45 epochs, this network has better results in deraining performance indicators compared than other deraining networks. In order to verify the object detection effectiveness of our network, ablation experiments are designed to perform object detection on deraining and denoising images by using YOLO v3 network, and the results show that the deraining and denoising network has improved the of object detection task in autonomous driving and enhances the robustness in the process of object detection.

## ACKNOWLEDGMENT

This research was supported by Information Center, Institute of Microelectronics of the Chinese Academy of Sciences and School of Automation, Beijing Institute of Technology.

## REFERENCES

[1] Hong Wang, Yichen Wu, Minghan Li, Qian Zhao, and Deyu Meng. A survey on rain removal from video and single image[J]. arXiv preprint arXiv:1909.08326, 2019.

[2] Wenhan Yang, Robby T. Tan, Jiashi Feng, Jiaying Liu, Zongming Guo and Shuicheng Yan. Deep join train detection and removal from a single image. In Proceedings of the IEEE Conference on Computer Vision and Pattern Recognition, 2017, pp. 1357‑1366.

[3] Wenhan,Yang; RobbyT,Tan; Jiashi,Feng; Zongming,Guo; Shuicheng, Yan and Jiaying,Liu. Joint rain detection and removal from a single image with contextualized deep networks. IEEE Transactions on Pattern Analysis and Machine Intelligence, vol. PP, no. 99, pp. 1‑1, 2019.

[4] Siyuan Li, Iago Breno Araujo, Wenqi Ren, Zhangyang Wang, and Eric K. Tokuda. Single image deraining: A comprehensive benchmark analysis. In Proceedings of the IEEE Conference on Computer Vision and Pattern Recognition, 2019, pp. 3838‑3847.

[5] Sachin Mehta, Mohammad Rastegari, Anat Caspi, Linda G. Shapiro, and Hannaneh Hajishirzi. Espnet. Efficient spatial pyramid of dilated convolutions for semantic segmentation. In Proceedings of the European Conference on Computer Vision (ECCV), pages 561‑580, 2018.

[6] N. Dalal and B. Triggs. Histograms of oriented gradients for human detection. In 2005 IEEE Computer Society Conference on Computer Vision and Pattern Recognition (CVPR), volume 1, pages 886‑893, 2005.

[7] Xiaozhi Chen, Kaustav Kundu, Ziyu Zhang, Huimin Ma, Sanja Fidler, and Raquel Urtasun. Monocular 3d object detection for autonomous driving. In 2016 IEEE Conference on Computer Vision and Pattern Recognition (CVPR), pages 2147‑2156, 2016.

[8] Taixiang Jiang, Tingzhu Huang, Xile Zhao, Liangjian Deng, and Yao Wang. A novel tensor-based video rain streaks removal approach via utilizing discriminatively intrinsic priors. In 2017 IEEE Conference on Computer Vision and Pattern Recognition (CVPR), pages 2818‑2827, 2017.

[9] Liwei Kang, Chiawen Lin, and Yuhsiang Fu. Automatic single-image-based rain streaks removal via image decomposition. IEEE transactions on image processing, 21(4):1742‑1755, 2011.

[10] Jin-Hwan Kim, Chul Lee, Jae-Young Sim, and Chang-Su Kim. Single-image deraining using an adaptive nonlocal means filter. In 2013 IEEE International Conference on Image Processing, pages 914‑917. IEEE, 2013.

[11] Duanyu Chen, Chiencheng Chen, and Liwei Kang. Visual depth guided color image rain streaks removal using sparse coding. IEEE transactions on circuits and systems for video technology, 24(8):1430‑1455, 2014.

[12] YuLi, Robby T Tan, Xiaojie Guo, Jiangbo Lu, andMichael SBrown. Rain streak removal using layer priors. In Proceedings of the IEEE conference on computer vision and pattern recognition, pages 2736‑2744, 2016.

[13] Wei Wei, Deyu Meng, Qian Zhao, Zongben Xu and Ying Wu. Semi-supervised transfer learning for image rain removal. In Proceedings of the IEEE Conference on Computer Vision and Pattern Recognition, pp. 3877‑3886, 2019.

[14] Xueyang Fu, Jiabin Huang, Xinghao Ding, Yinghao Liao, and John Paisley. Clearing the skies: A deep network architecture for single-image rain removal. IEEE Transactions on Image Processing, 26(6):2944‑2956, 2017.

[15] Xueyang Fu, Jiabin Huang, Delu Zeng, Yue Huang, Xinghao Ding, and John Paisley. Removing rain from single images via a deep detail network. In Proceedings of the IEEE Conference on Computer Vision and Pattern Recognition, pages 3855‑3863, 2017.

[16] K. Garg and S.K. Nayar. Detection and removal of rain from videos. In Proceedings of the 2004 IEEE Computer Society Conference on Computer Vision and Pattern Recognition, 2004. CVPR 2004., volume 1, pages 528‑535, 2004.

[17] Kshitiz Garg and Shree K Nayar. Vision and rain. International Journal of Computer Vision, 75(1):3‑27, 2007.

[18] He Zhang, Vishwanath Sindagi, and Vishal M. Patel. Image deraining using a conditional generative adversarial network. IEEE Transactions on Circuits and Systems for Video Technology, 2017.

[19] Wei Wei, Deyu Meng, Qian Zhao, Zongben Xu, and Ying Wu. Semi-supervised transfer learning for image rain removal. In Proceedings of the IEEE/CVF Conference on Computer Vision and Pattern Recognition (CVPR), June 2019.

[20] Zongsheng Yue, Jianwen Xie, Qian Zhao, Deyu Meng, Semi-Supervised Video Deraining With Dynamical Rain Generator. In Proceedings of the IEEE/CVF Conference on Computer Vision and Pattern Recognition (CVPR), 2021, pp. 642-652.

[21] Kshitiz Grag, Shree K Nayar. Photometric model of a rain drop[J]. CMU Technical Report, 2003.